\documentclass{czjp}
\usepackage{epsfig}  
\begin{document}
\begin{frontmatter}
\title{RESULTS ON THE NUCLEON SPIN STRUCTURE}
\author{Eva-Maria~Kabu{\ss}\thanks{supported by the BMBF}}
\address{Inst. f\"ur Kernphysik, Mainz University, D-55099 Mainz, Germany}
\author{for the Spin Muon Collaboration} %
\runningauthor{Eva-Maria~Kabu{\ss} } 
\runningtitle{Nucleon spin structure }
\begin{abstract}
SMC performed an investigation of the spin structure of the nucleon
by measuring deep inelastic scattering of polarised muons off polarised
protons and deuterons. Asymmetries and spin structure functions were obtained 
for
$x>0.0008$ and $Q^2>0.2$~GeV$^2$. Using a next-to-leading order QCD analysis
of all experimental results 
polarised parton distributions and their first moments were determined.
All data show a clear violation of the Ellis-Jaffe sum rule. The Bjorken
sum rule is found to be valid and is tested to the 10\% level.
\end{abstract}
%
%
\end{frontmatter}
\section{Introduction}
To investigate the spin structure of the nucleon SMC studies deep inelastic
scattering of polarised muons off polarised protons and deuterons. 
From the measured longitudinal asymmetries the spin structure functions, $g_1$, can be
extracted by
\begin{equation}
A_{\parallel}=\frac{d\sigma^{\uparrow \downarrow}-d\sigma^
{\uparrow \uparrow}}{d\sigma^{\uparrow \downarrow}+d\sigma^{\uparrow 
\uparrow}} \approx D \frac{g_1}{F_1}.
\label{emk-apar}
\end{equation}
The depolarisation factor $D$ gives the polarisation transfer from the lepton
to the nucleon and increases with the energy transfer. $F_1$ is the unpolarised
structure function.

In the quark parton modell (QPM) $g_1$ is related to
the quark helicity distributions functions, 
$\Delta q=q^{\uparrow}-q^{\downarrow}$, by
\begin{equation}
g_1(x,Q^2)=\frac{1}{2}\sum\limits_{\rm q} e_{\rm q}^2 (\Delta q(x,Q^2) + 
\Delta \bar{q}(x,Q^2))
\end{equation}
where $Q^2$ is the negative square of the four momentum transfer, Bjorken $x$ the
momentum fraction carried by the struck quark and  $e_{\rm q}$ the quark
charge. The notation $q^{\uparrow}$ ($q^{\downarrow}$) refers to parallel
(antiparallel) orientation of the quark and the nucleon spins.
The first moments 
$a_{\rm q}(Q^2)=\int\limits_{0}^{1} (\Delta q(x,Q^2)+\Delta \bar{q}(x,Q^2))
{\rm d}x$ and
$\Gamma_1(Q^2)=\int\limits_{0}^{1} g_1(x,Q^2){\rm d}x$ can be used to
study the contribution of quark helicities to the nucleon spin and to test the
Ellis-Jaffe \cite{emk-ej} and Bjorken \cite{emk-bj} sum rules.

The first measurement of the Ellis-Jaffe sum rule for the first moment of 
$g_1^p$
by the EMC \cite{emk-emc} showed a strong violation of the sum rule.
Interpreted in the QPM this  pointed to a 
very small contribution of quark helicities to the proton spin and a negative
polarisation of the strange sea quarks. This led to a series of new
experiments at CERN, SLAC and also at DESY studying the spin structure of the 
proton and the neutron in more detail.
These new data also allow to test the fundamental Bjorken sum rule  
$\Gamma_1^{\rm p}-\Gamma_1^{\rm n} =\left| \frac{g_{\rm A}}{g_{\rm V}}
\right|/6$. In addition, the 
structure function data are precise enough to perform
next-to-leading order perturbative QCD analyses and determine the
polarised parton distributions.

\section{The experiment}
The SMC experiment (NA37) was located at the M2 muon beam line of the CERN SPS
and took data using polarised proton and deuteron targets from 1992 to 1996.

The 190 and 100 GeV muon beam with about $4 \cdot 10^7 \mu$ per 2.2~s spill was
naturally polarised. The polarisation
was determined by two independent measurements using a dedicated second
spectrometer downstream of the main experiment. The first method measured the 
energy spectrum
of the decay electrons from muon decay, the second one used the 
cross section asymmetry in polarised muon electron scattering \cite{emk-pol}. 
The results were  $p_{\mu}=-0.795 \pm 0.019$ at
187.4~GeV and  $p_{\mu}=-0.81 \pm 0.03$ at 99.4~GeV.

The heart of the experiment was the polarised solid state target. It was
polarised using dynamic nuclear polarisation in a 2.5~T longitudinal field of a
superconducting magnet. The target temperature was maintained at 
$\approx 50-300$~mK
during normal operation by a dilution refrigerator. From 1992 to 1995 SMC used
butanol or deuterated butanol as target material whereas for the 1996 data
taking ammonia was chosen improving the dilution factor $f$ by 30\% from about
0.13 for butanol to about 0.17 for ammonia. 
The target
was divided into two cells of 65~cm length polarised in opposite
directions. To reduce systematic uncertainties due to changes in the
detector response the target polarisation was reversed every 5~hours by
rotating the magnetic field with the help of an additional 0.5~T dipole 
magnet. The average longitudinal polarisation was
$p_{\rm T}=\pm 0.90$ for protons and $\pm50$\% for 
deuterons. It was measured to an accuracy of 2-3\% using 9 NMR
coils embedded in the target material. In addition to the continuous
measurement of the proton polarisation the polarisation of $^{14}N$ was
determined in dedicated measurements \cite{emk-nitr} and found to depend on the proton
polarisation as predicted by the equilibrium spin temperature relation. 
Typical nitrogen polarisations were $\pm 0.14$ for proton polarisations of 
$\pm 0.89$.

The scattered muons and the produced forward going hadrons were detected in an
open forward spectrometer using about 150 wire chamber planes placed upstream
and downstream of a 
large gap
dipol magnet. Due to the large number of wire planes the measured asymmetries 
were hardly sensitive to occasional problems with single planes. There was no
particle identification for hadrons, but electron hadron separation was
obtained with the help of a lead/iron scintillator calorimeter.

The measured counting rates for longitudinal beam and target
polarisations were used to determine the raw asymmetry
\begin{equation}
A^{\rm raw}_{\parallel}(x,Q^{2}) = \frac{N^{\uparrow \downarrow}-
N^{\uparrow \uparrow}}{N^{\uparrow \downarrow}+N^{\uparrow \uparrow}},
\end{equation}
which is related to the lepton nucleon ($A_{\parallel}$) and photon nucleon 
asymmetries ($A_1$, $A_2$) by
\begin{equation}
A^{\rm raw}_{\parallel}=p_{\mu}p_{\rm T}f A_{\parallel}=
p_{\mu}p_{\rm T}fD(A_{1}+\eta A_{2}).
\label{emk-raw}
\end{equation}
In a separate measurement using transversly polarised targets the transverse
asymmetry 
\begin{equation}
A^{\rm raw}_{\perp}=\frac{N^{\uparrow \rightarrow}-
N^{\uparrow \leftarrow}}{N^{\uparrow \rightarrow}+N^{\uparrow \leftarrow}}
=p_{\mu}p_{\rm T}fd(A_{2}+\eta' A_{1})
\end{equation}
was measured. $D$, $d$, $\eta$ and $\eta'$ are known kinematic factors.
$A_{2}$ was found to be small in the kinematic range of the SMC data 
\cite{emk-a2,emk-a1long} and thus the second term in eq.~\ref{emk-raw} can be neglected.

\section{Results for {\boldmath $A_1$} and {\boldmath $g_1$}}
\begin{figure}[htb]
\begin{center}
\mbox{
\epsfig{file=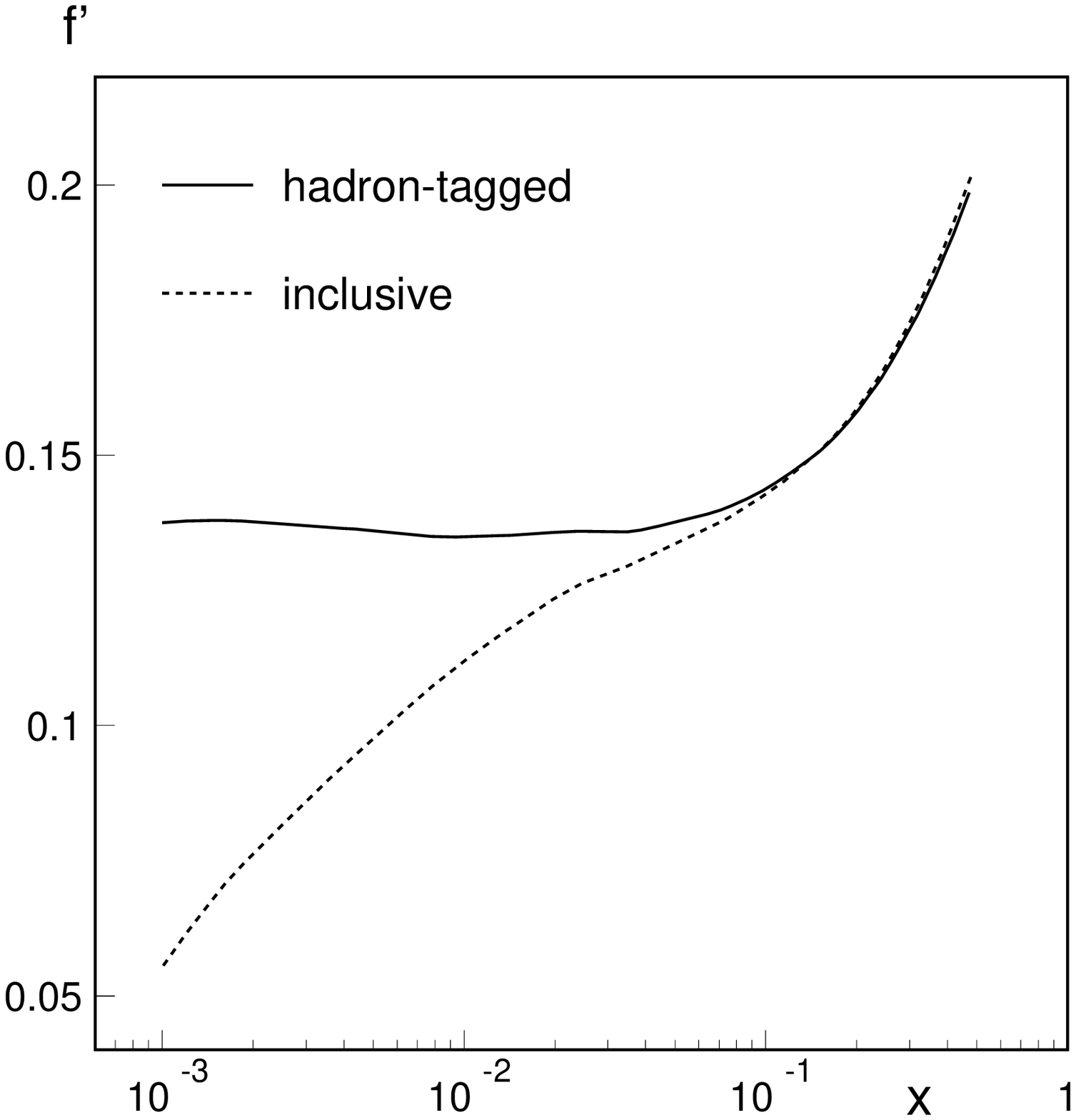,width=0.37\textwidth}
\hspace{0.8cm}
\epsfig{file=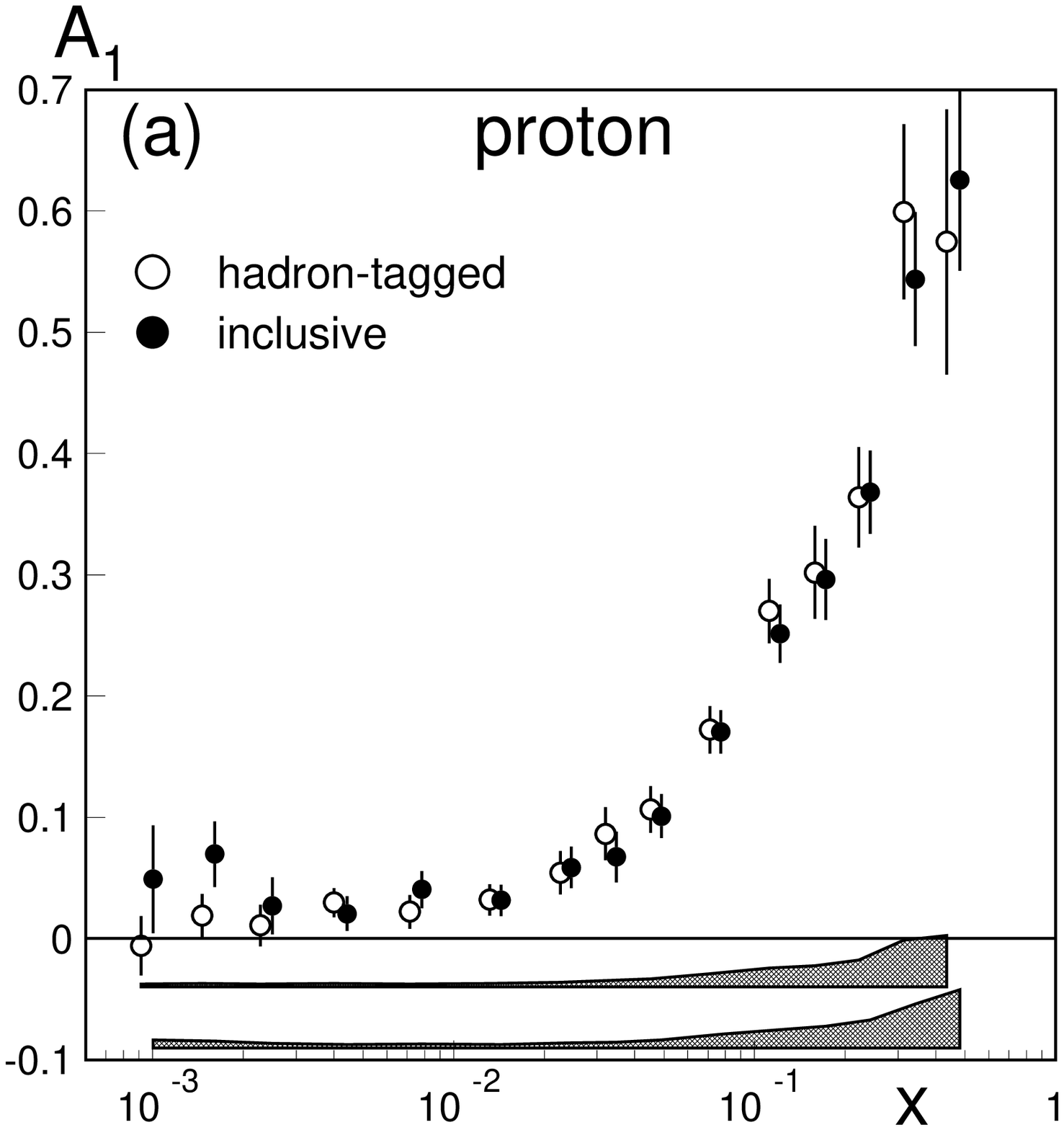,width=0.46\textwidth}
}
\caption{\label{emk-dilfac} Comparison a) of the effective dilution factor $f'$, b) of the
proton asymmetries $A_1^{\rm p}$ from the standard analysis method and the hadron
method.
}
\end{center}
\end{figure}

Inclusive asymmetries are determined using two methods. In the
first one, called standard method, only scattered muons are used, in the
second one, any hadron in addition to the scattered muon signals a true
deep inelastic scattering and provides a discrimination between these events 
and background events like radiative events or elastic electron muon
scattering \cite{emk-asym98}. 
Thus, the effective dilution factor 
$f'=f \cdot \sigma^{1\gamma}/\sigma^{\rm tot}$ is much larger for
the hadron method than for the standard one (see fig.~\ref{emk-dilfac}a ).

\begin{figure}[htb]
\begin{center}
\mbox{
\epsfig{file=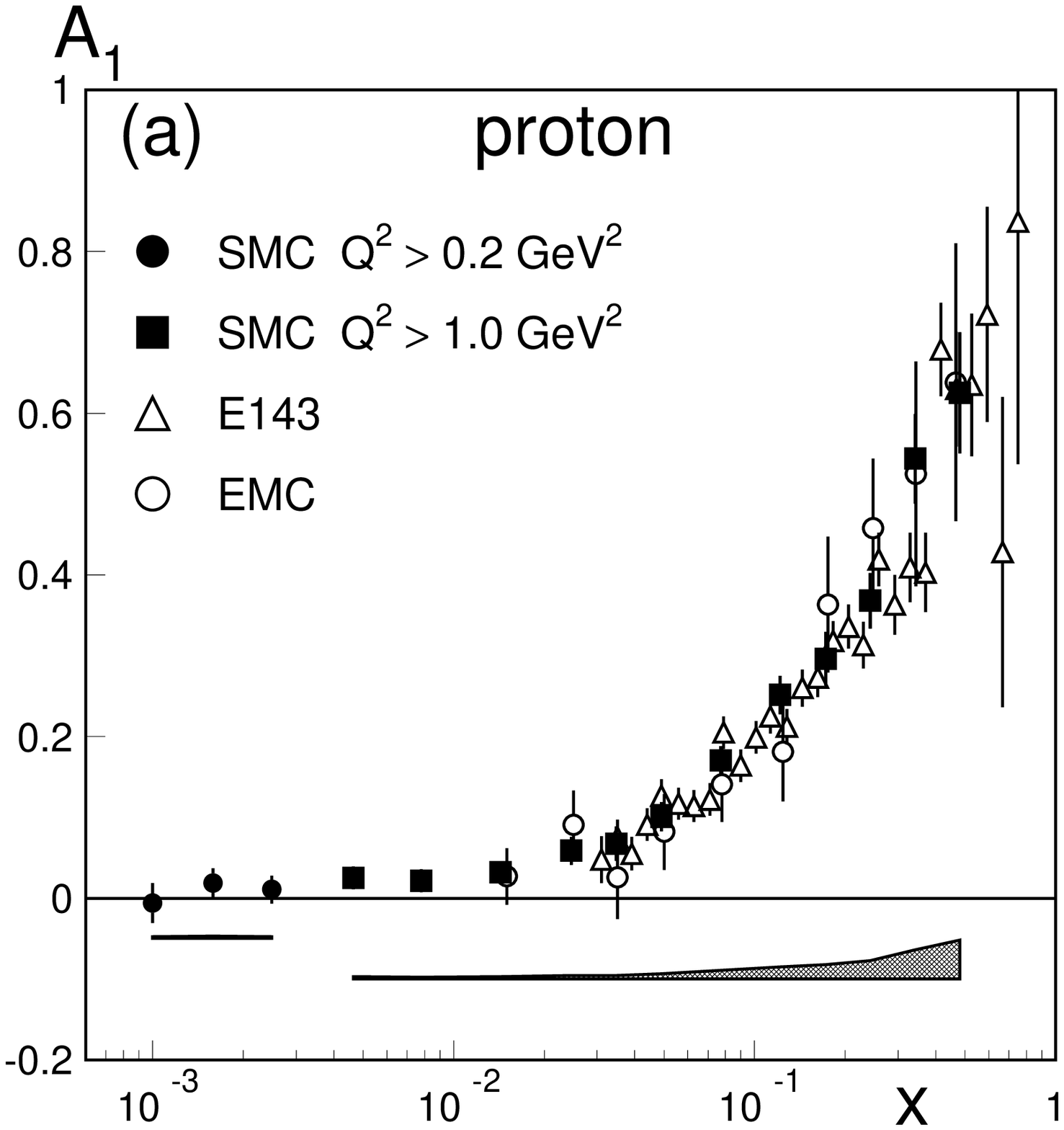,width=0.45\textwidth}
\epsfig{file=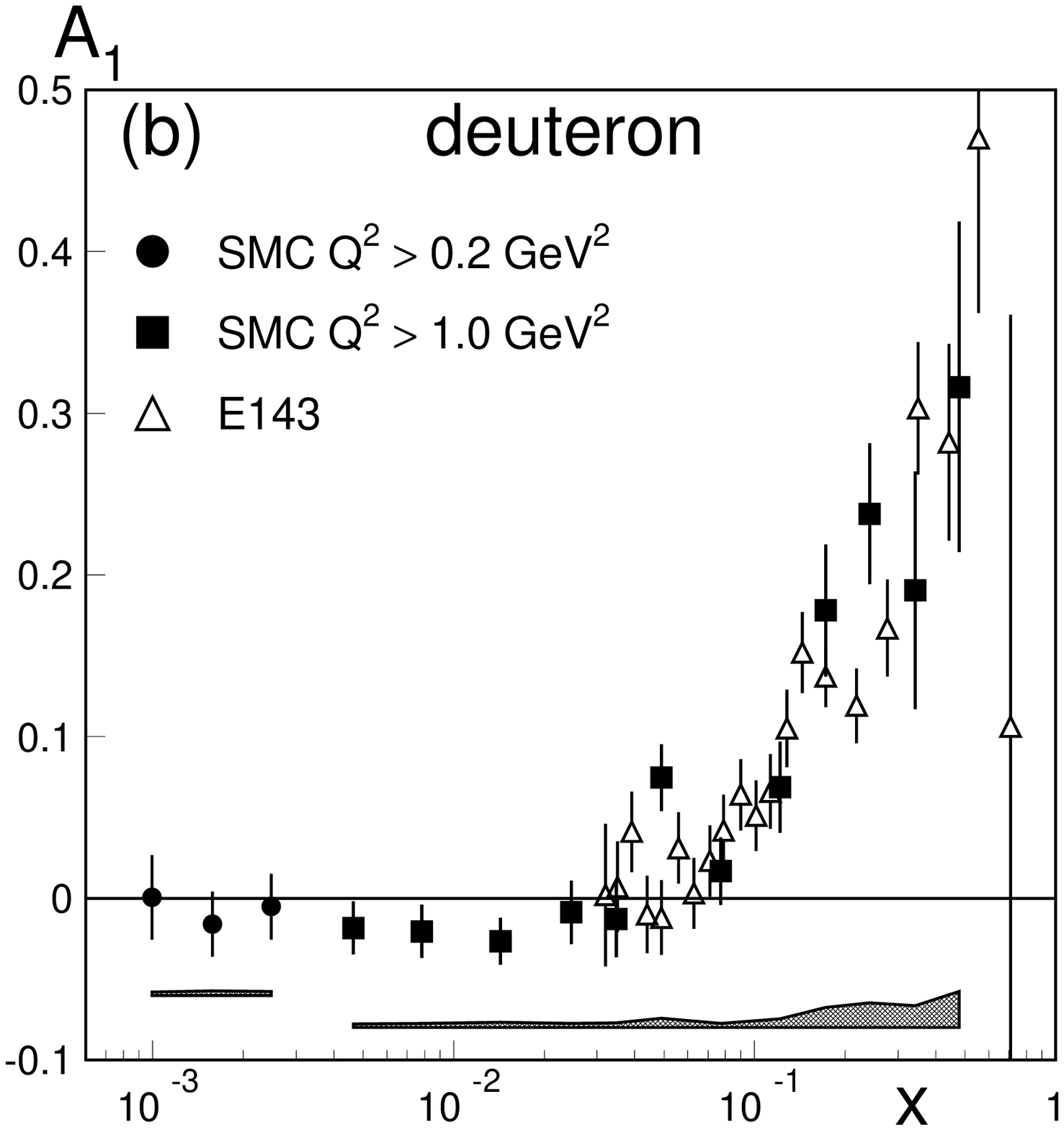,width=0.45\textwidth}
}
\caption{\label{emk-a1} The optimal data set of SMC for $A_1$ together with results
from other experiments. Statistical errors are shown as error bars, while the
shaded bands below indicate the systematic uncertainty for the SMC measurement.}
\end{center}
\end{figure}

In the SMC spectrometer mainly forward produced charged hadrons with momenta
$p_{\rm h}>5$~GeV were detected. Events with a high mass $W$ of the final
state contain in general several hadrons with high momenta. High $W$ events
are typically found at small $x$, so that a good acceptance for deep inelastic
events is obtained for $x<0.1$, whereas in the high $x$ region many hadrons
are not detected due to the low momenta at low $W$. In addition cuts are
applied to reject electrons from conversion of Bremstrahlung photons.
Monte Carlo studies show
that more than 80\% of the deep inelastic events are found by requiring at
least one additional hadron to the scattered muon. 
The results for $A_1^{\rm p}$ from both methods are compared in fig.~\ref{emk-dilfac}b for $x>0.0008$
and $Q^2>0.2$~GeV$^2$. Good agreement is obtained between the two results
with the statistical erros bars smaller for $x<0.1$ for the hadron method.
The optimal SMC data set for $A_1^{\rm p}$ and $A_1^{\rm d}$ is shown in 
fig.~\ref{emk-a1} in comparison to measurements from the EMC \cite{emk-emc} at a similar $Q^2$
and the lower $Q^2$~data from E143 \cite{emk-e143}.
Systematic uncertainties are shown as bands at the bottom.
At low $x$, the main contributions are due to the uncertainty of radiative 
corrections, the neglect of $A_2$ and acceptance variations.
At high $x$ they stem from
uncertainties of $R$ and the beam and target polarisation.  

In addition to the $x$ dependence also the $Q^2$ dependence of $A_1$ was
investigated. No $Q^2$ dependence is observed in the kinematic range of the 
SMC data.

\begin{figure}[htb]
\hspace{-0.4cm}
\epsfig{file=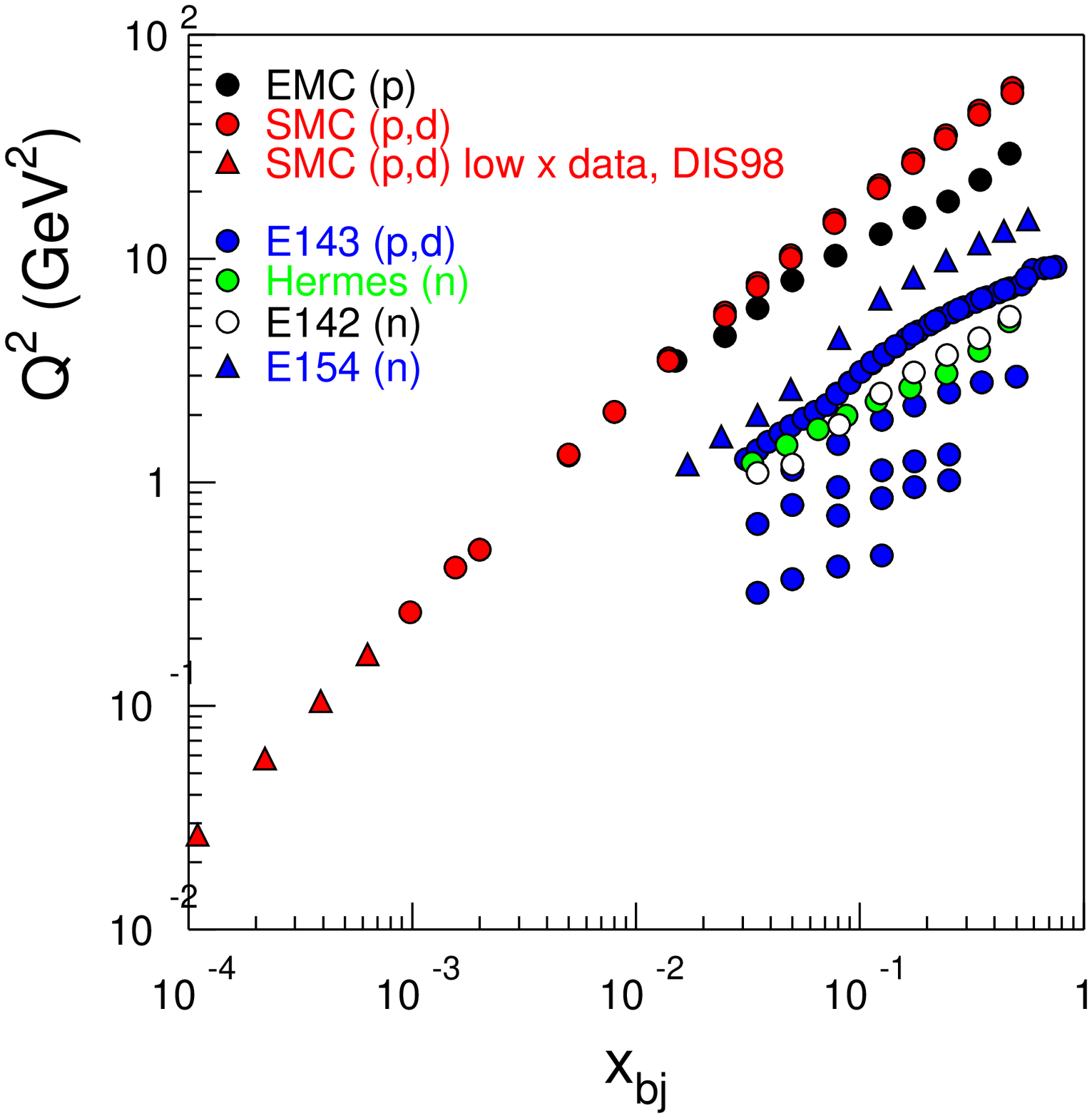,width=0.5\textwidth}
\hspace{0.5cm}
\epsfig{file=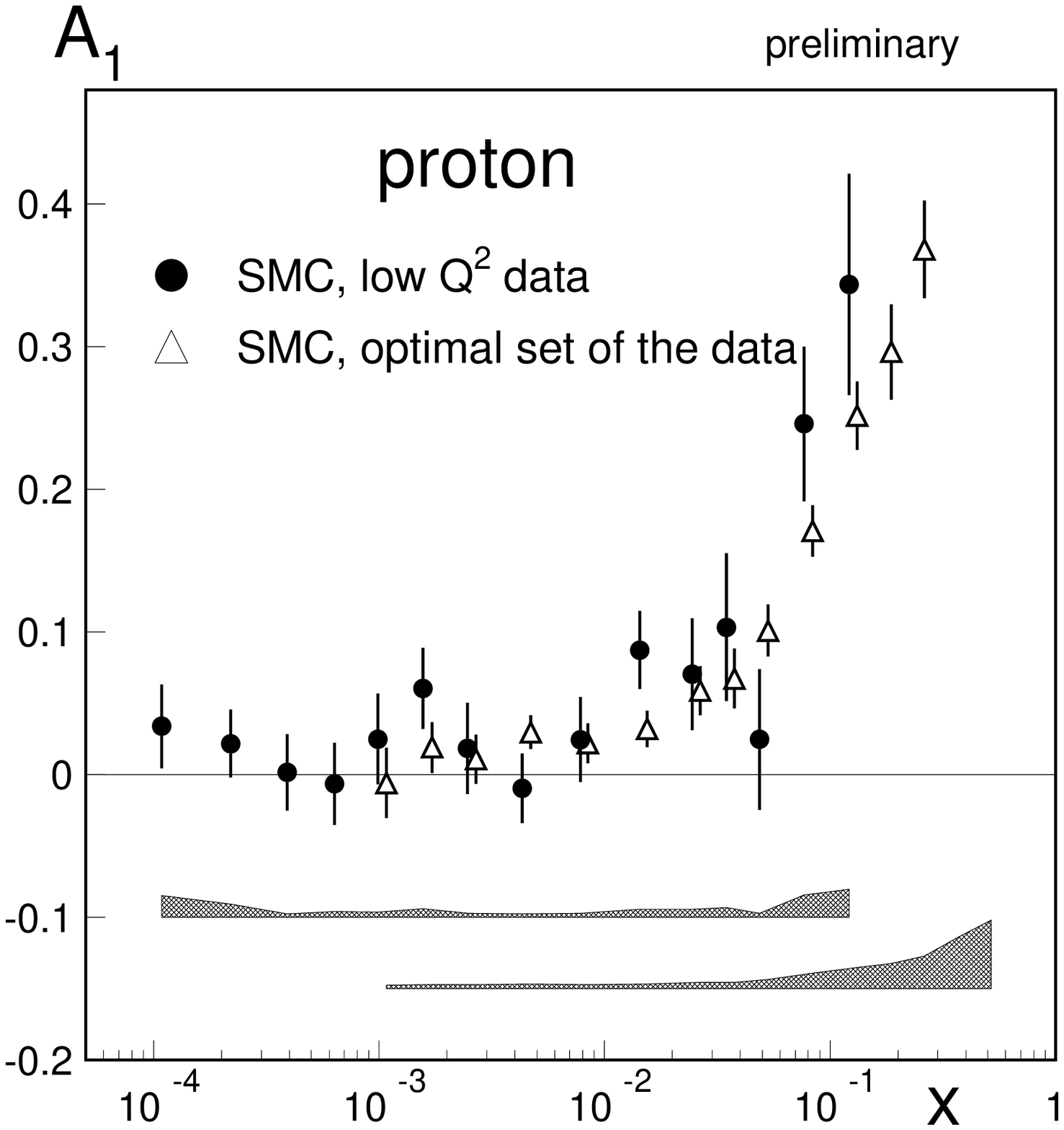,width=.52\textwidth}
\caption{\label{emk-lowx} a) Kinematic acceptance of the "low $x$" trigger compared to other
experiments, b) preliminary results on $A_1^{\rm p}$ compared to the SMC 
optimal data set. The error bars are statistical errors, the bands indicate
the
systematic uncertainties.}
\end{figure}

An extension of the measured asymmetries down to $x=6\cdot 10^{-5}$ with
$Q^2>0.01$~GeV$^2$  is under study.
In this kinematic region the inclusive events are dominated by muon scattering
off atomic electrons. In 1995/6 data were collected using a dedicated "low $x$" 
trigger in which both a minimal energy deposit in the hadronic part of the 
calorimeter and the detection of a muon was demanded \cite{emk-a1lowx}.
Fig.~\ref{emk-lowx}a shows the kinematic acceptance of this trigger compared
to other experiments, while the preliminary results for $A_1^{\rm p}$ are 
shown in fig.~\ref{emk-lowx}b. 
The two SMC data sets are compatible in the overlap region.
At low $x$ no significant spin effects are visible.

\begin{figure}[htb]
\begin{center}
\mbox{
\epsfig{file=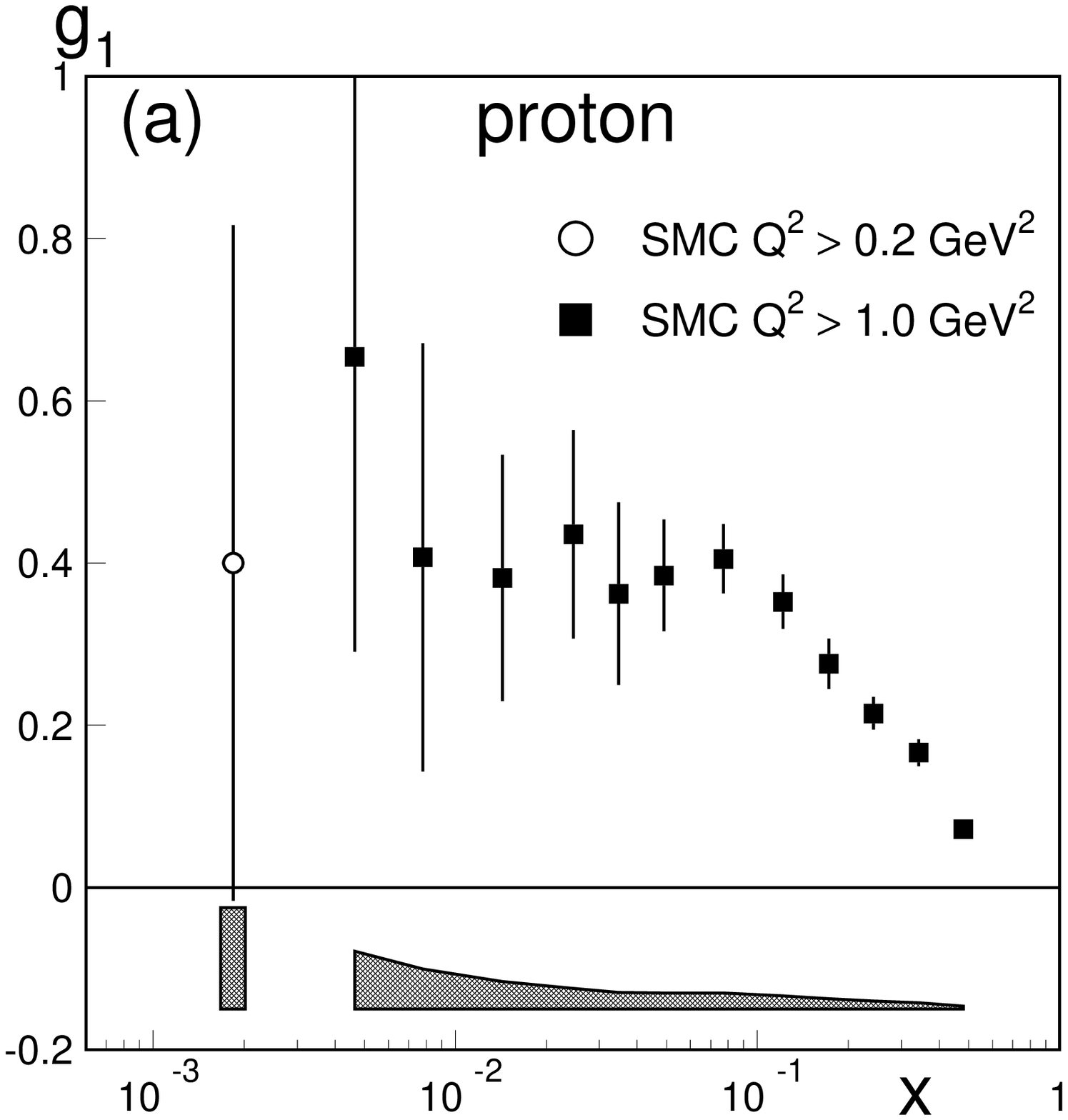,width=.45\textwidth}
\epsfig{file=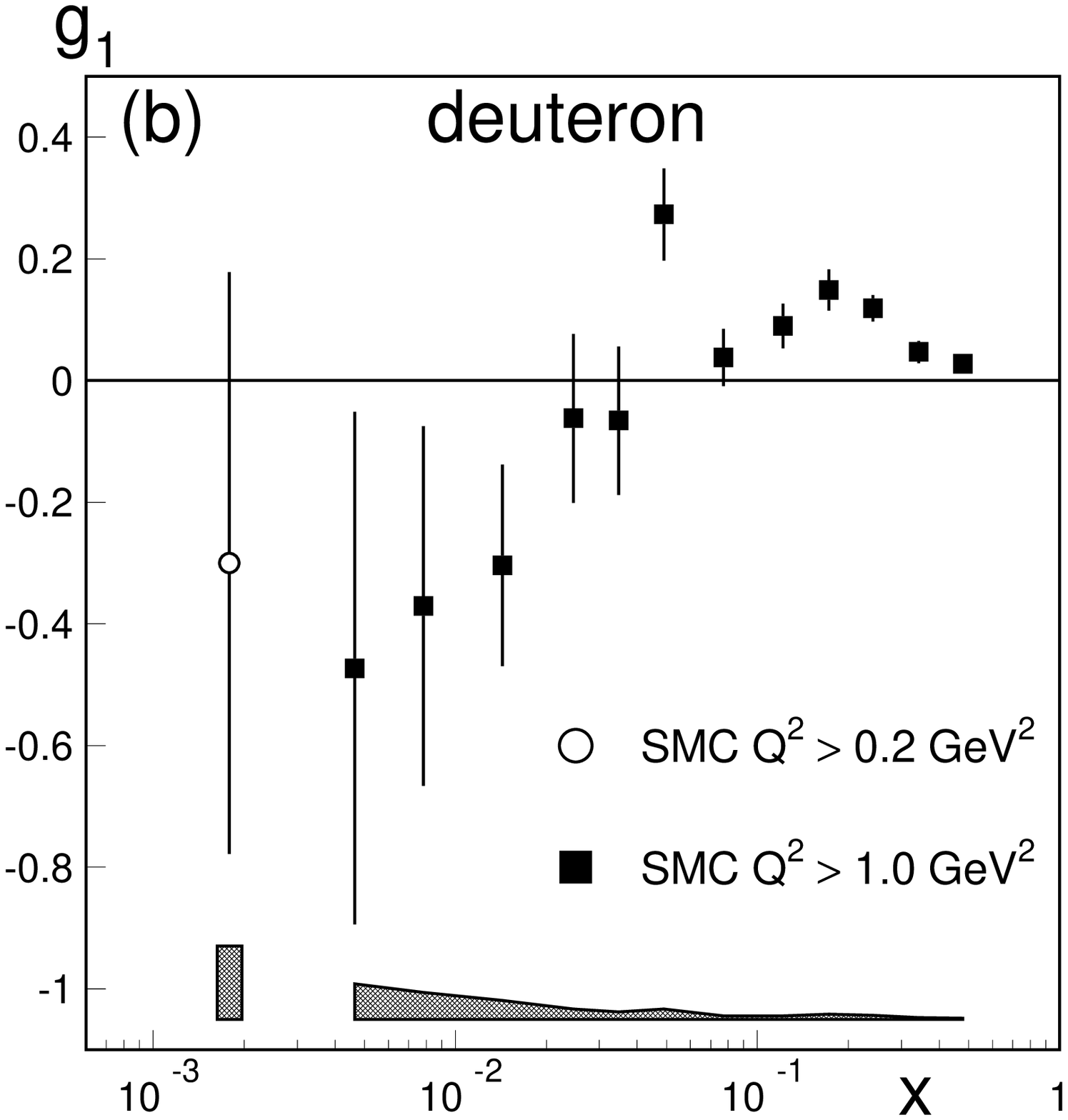,width=.45\textwidth}
}
\caption{\label{emk-g1} a) $g_1^{\rm p}$ and b) $g_1^{\rm d}$ vs. $x$ at the measured $Q^2$ for
  the
optimal data set. The statistical uncertainties are shown as error bars, and the
bands indicate the size of the systematic uncertainties.}
\end{center}
\end{figure}

For $Q>0.2$~GeV$^2$ the spin structure functions, $g_1$, are calculated 
from the measured 
asymmetries with eq.~\ref{emk-apar} using $F_1=F_2/2x(1+R)$, where $F_2(x,Q^2)$
and $R(x,Q^2)$ are taken from unpolarised measurements (see fig.~\ref{emk-g1}).

\section{QCD analysis}

\begin{figure}[htbp]
\begin{center}
\mbox{
\epsfig{file=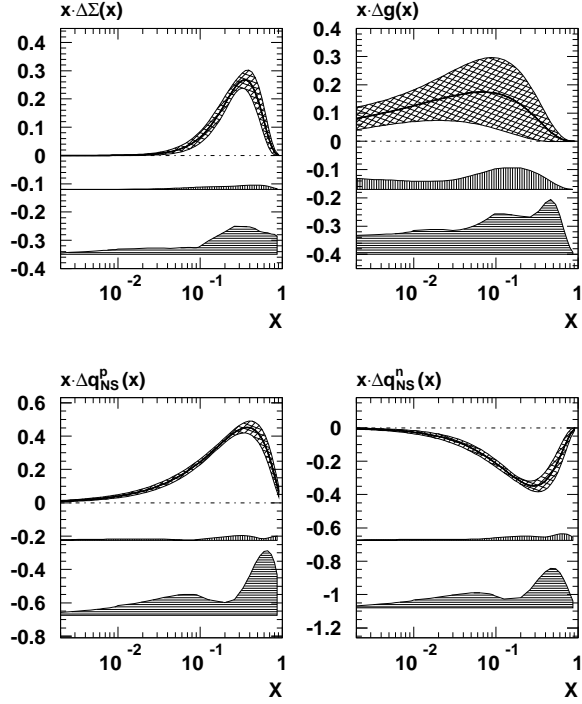,width=.7\textwidth}
}
\caption{\label{emk-pdf}  Polarised parton distributions determined at 
$Q^2=1$~GeV$^2$. The uncertainties are statistical (cross hatched band),
  experimental  systematic (vertically hatched band), and
  theoretical (horizontally hatched band).
}
\end{center}
\end{figure}

To determine the first moment, $\Gamma_1(Q_0^2)$, the results for $g_1(x,Q^2)$
at a fixed $Q^2=Q_0^2$ are needed. Though the precision of the available  data
and the $Q^2$ range do not allow a direct measurement of a possible $Q^2$
dependence of $A_1=g_1/F_1$ different $Q^2$ behaviours of $g_1$ and $F_1$ are
expected from perturbative QCD. Therefore, the $Q^2$ dependence of $g_1$ is
estimated from a next-to-leading-order (NLO) 
perturbative QCD analysis \cite{emk-qcd98} of all available proton,
neutron and deuteron data with $Q^2>1$~GeV$^2$. 
In this analysis $g_1$ is decomposed into
polarised quark singlet $\Delta \Sigma$, gluon $\Delta g$ and nonsinglet 
distributions $\Delta q^{\rm{NS}}$ for the proton
and the neutron. 
The NLO analysis is done in the Adler-Bardeen  
and the $\overline{\mbox{MS}}$ scheme \cite{emk-bfr}.
In the fit some parameters like the normalisation of the nonsinglet 
distributions were constrained by using the neutron and hyperon $\beta$ decay
constants assuming SU(3) flavour. The results for the polarised parton
distributions  $x\Delta \Sigma$, $x\Delta g$ and $x\Delta q^{\rm{NS}}$ are
shown in fig.~\ref{emk-pdf}. 

In the QPM the quark contribution to the nucleon spin is given by the singlet
axial charge 
$a_0=a_{\rm u}+a_{\rm d}+a_{\rm s}$ whereas it is ambiguous
in NLO pQCD due to the axial anomaly. Thus, the first moment of $\Delta \Sigma$,
$a_{\Sigma}$
is equal to $a_0$ in the $\overline{\mbox{MS}}$ scheme, while in the AB scheme 
the gluon contribution must be subtracted.
The results 
$a_0^{\rm AB}=0.23\pm 0.07(\mbox{sta})\pm0.19(\mbox{sys} \& \mbox{th})$ and 
$a_0^{\rm \overline{\rm {MS}}}=0.19\pm 0.05(\mbox{sta})\pm0.04
(\mbox{sys} \& \mbox{th})$
are similar in both schemes and correspond to about 1/3 of the naive Ellis-Jaffe
expectation  of 0.58.
The result for the gluon contribution 
$a_{\rm g}(Q^2)=\int\limits_{0}^{1} \Delta g(x,Q^2){\rm d}x$
\begin{equation}
a_{\rm g}^{\rm AB}=0.99\;^{+1.17}_{-0.31}\; ({\rm sta})\;\; ^{+0.42}_{-0.22}\;
                             ({\rm sys})\;\; ^{+1.43}_{-0.45}\;
                             ({\rm th})
\end{equation}
shows that very little can be said about this quantity on the basis of the 
present data.

\section{Sum rules}

\begin{figure}[htb]
\begin{center}
\mbox{
\epsfig{file=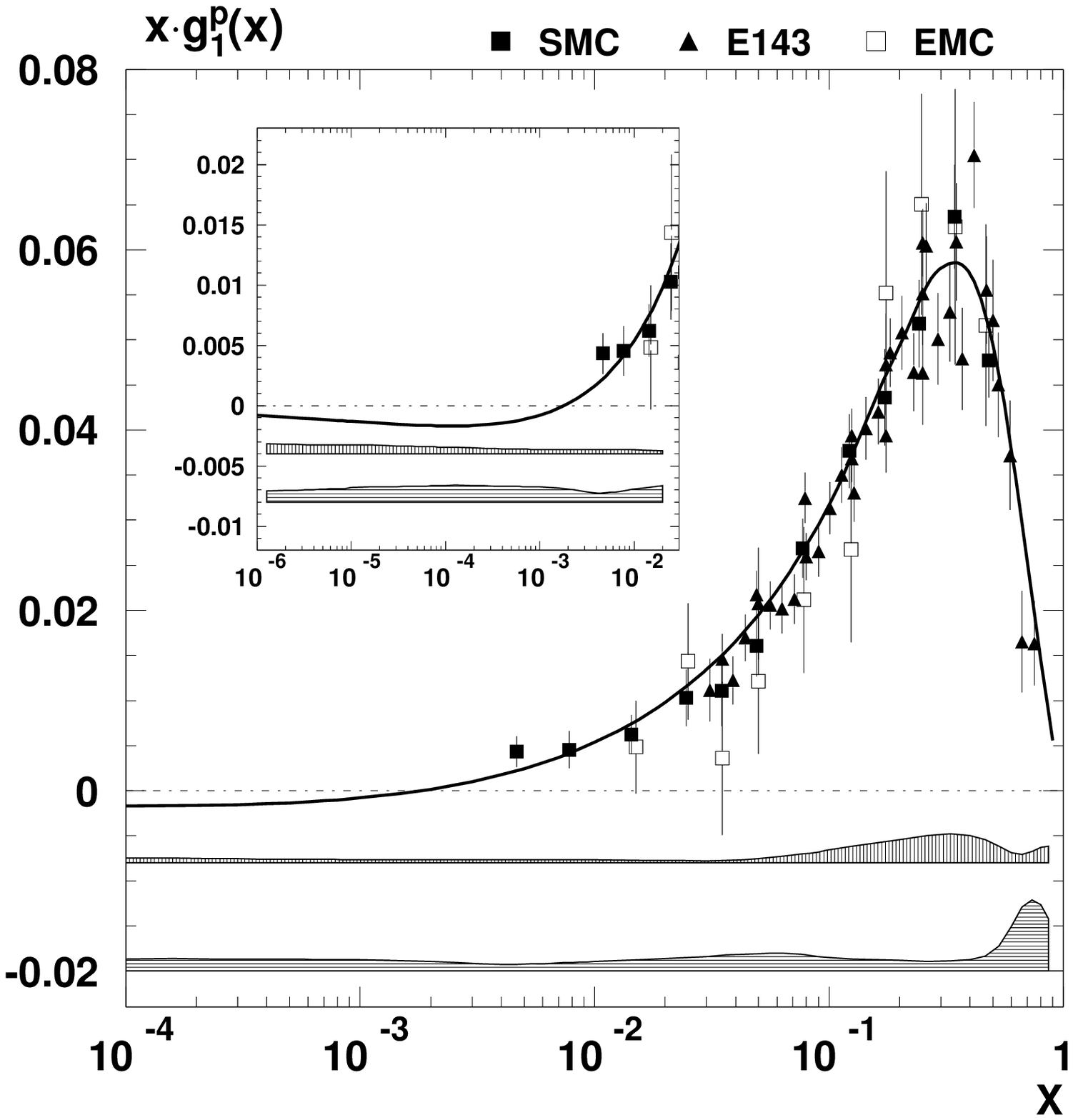,width=0.5\textwidth}
\epsfig{file=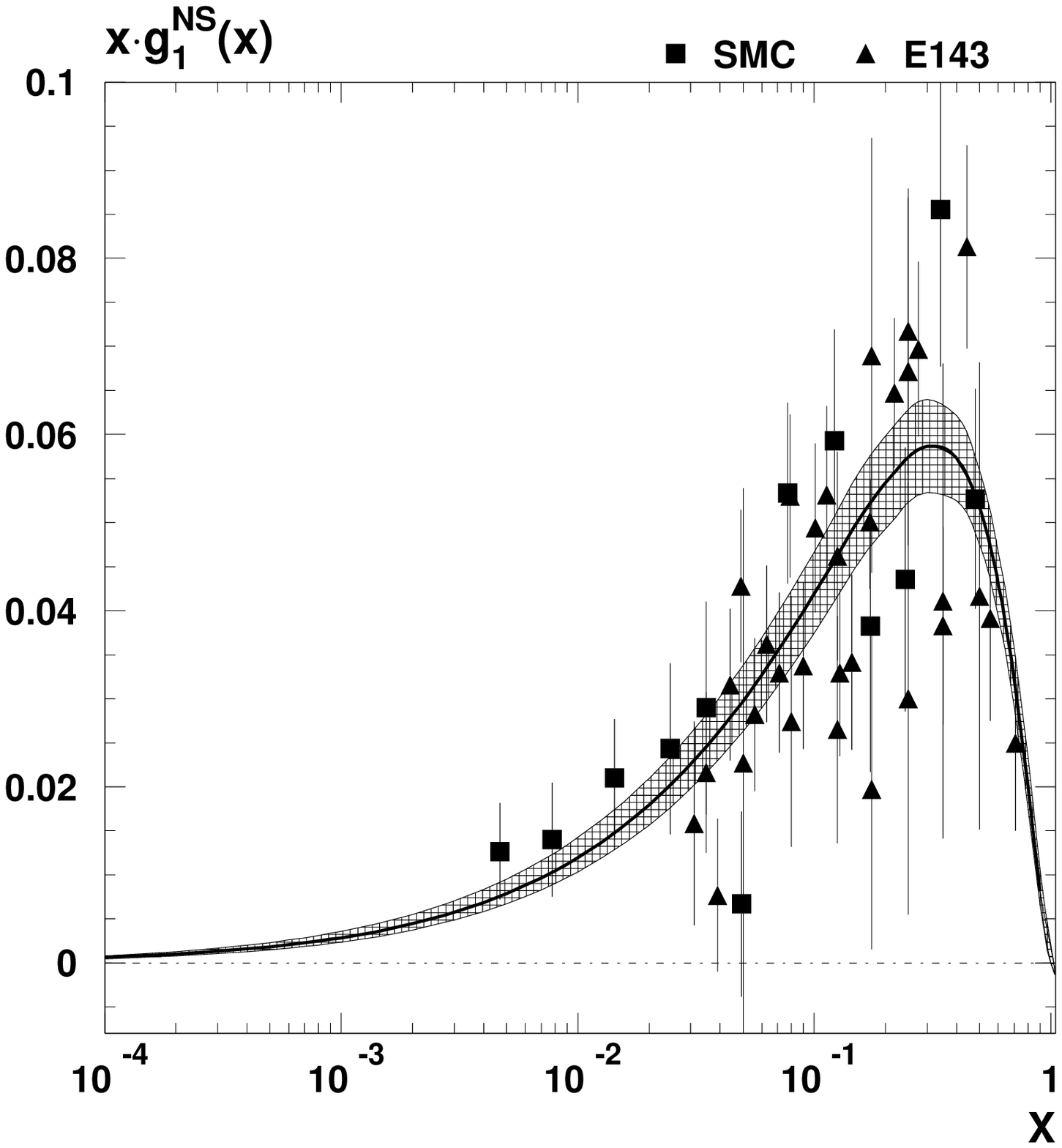,width=0.5\textwidth}
}
\caption{\label{emk-sr}
a) $xg_1^{\rm p}$ vs. $x$ for the world data with the
uncertainties of the fit due to experimental 
systematics and theoretical sources shown by the bands, 
b) $xg_1^{\rm NS}$ vs. $x$ from 
E143 and
SMC with the QCD fit at $Q^2=5$~GeV$^2$ with the error band representing the 
systematic and theoretical uncertainty of the fit. 
All points are shown with their statistical errors.}
\end{center}
\end{figure}

Using the results of the QCD analysis $g_1(x,Q_0^2=5$~GeV$^2)$ is
obtained from
$g_{1}(x,Q_0^2) =
g_{1}(x,Q^{2})
    + \left[g_{1}^{\rm fit}(x,Q_0^2) - g_{1}^{\rm fit}(x,Q^{2}) \right]$.
The extrapolation of $g_1$ to $x=0$ and
$x=1$ is done by integrating the fit results for $x<0.003$ and $x>0.8$
(see fig.~\ref{emk-sr}a). The results at $Q^{2}_{0}=5$~GeV$^{2}$
\begin{eqnarray}
\Gamma^{\rm p}_1 &=&     0.121 \pm 0.003 \; ({\rm sta})\;\;\pm 0.005 \; 
({\rm sys})\;\;\pm 0.017\; ({\rm th})\;\; \nonumber  \\
\Gamma^{\rm d}_1 &=&     0.021 \pm 0.004 \; ({\rm sta})\;\;\pm 0.003 \; 
({\rm sys})\;\;\pm 0.016\; ({\rm th})\;\; \nonumber  \\
\Gamma^{\rm n}_1 &=&   -0.075 \pm 0.007 \; ({\rm sta})\;\;\pm 0.005 \; 
({\rm sys})\;\;\pm 0.019\; ({\rm th})\;\;  
\end{eqnarray}
show a violation of the Elle-Jaffe sum rules of about 3 $\sigma$.

To test the Bjorken sum rule the fit has to be repeated releasing the
constraints for the nonsinglet moments using $g_{\rm{A}}/g_{\rm{V}}$ as
a free parameter. A consistent value of $g_{\rm {A}}/g_{\rm{V}}$ with the
nominal value used above is obtained. The resulting value for the Bjorken
sum rule at $Q^2=5$~GeV$^2$ 
\begin{equation}
\Gamma_1^{\rm p}-\Gamma_1^{\rm n}
   \ = \ 0.174\;\; ^{+0.024} _{-0.012}
\end{equation}
is in excellent agreement with the theoretical value of 
$ \Gamma_1^{\rm p} - \Gamma_1^{\rm n} = 0.181 \pm 0.003$.
An alternative way to determine $ \Gamma_1^{\rm p} - \Gamma_1^{\rm n}$
restricts the QCD analysis to the nonsinglet distribution which is decoupled
from the evolution of $\Delta \Sigma$ and $\Delta g$. Here, data points 
for $g_1^{\rm {p}}$ and $g_1^{\rm {n}}$ at the same kinematics are needed
so that only the SLAC E143 and SMC can be used presently
(see fig.~\ref{emk-sr}b)). 
The result
\begin{equation}
\Gamma_1^{\rm p} - \Gamma_1^{\rm n}
   \ = \ 0.181\;\; ^{+0.026} _{-0.021}
\end{equation}
is constistent with the above result, but due to the smaller data set the 
experimental errors are larger. The theoretical error, however, is smaller
than for the first method, so that with the forthcoming data from E155 and HERMES
a significant improvement is expected.


\end{document}